\documentclass[aps,preprint,prl,amsmath,superscriptaddress,floatfix,showpacs,showkeys]{revtex4-1}
\usepackage{graphicx}
\usepackage{hyperref}

\begin{document}

\title{Photonic hyperuniform networks by silicon double inversion of polymer templates}

\author{Nicolas Muller}
\affiliation{Department of Physics, University of Fribourg, CH-1700 Fribourg, Switzerland}

\author{Jakub Haberko}
\affiliation{Faculty of Physics and Applied Computer Science, AGH University of Science and Technology,  al. Mickiewicza 30, 30-059 Krakow, Poland}

\author{Catherine Marichy}
\affiliation{Universit\'e de Lyon, Laboratoire des Multimat\'{e}riaux et Interfaces, Villeurbanne Cedex, France, 69622}

\author{Frank Scheffold}
\email{Frank.Scheffold@unifr.ch}  
\affiliation{Department of Physics, University of Fribourg, CH-1700 Fribourg, Switzerland}

\begin{abstract}
Hyperuniform disordered networks belong to a peculiar class of structured materials predicted to possess partial and complete photonic bandgaps for relatively moderate refractive index contrasts. The practical realization of such photonic designer materials is challenging however, as it requires control over a multi-step fabcrication process on optical length scales. Here we report the direct-laser writing of hyperuniform polymeric templates followed by a silicon double inversion procedure leading to high quality network structures made of polycrystalline silicon. We observe a pronounced gap in the shortwave infrared centered at a wavelength of $\lambda_{\text{Gap}}\simeq $ 2.5 $\mu$m, in nearly quantitative agreement with numerical simulations.  In the experiments the typical structural length scale of the seed pattern can be varied between  2 $\mu$m and 1.54 $\mu$m leading to a blue-shift of the gap accompanied by an increase of the silicon volume filling fraction.
\end{abstract}

\maketitle

\section{Introduction}
Photonic crystal structures have drawn a lot of attention over the last two decades but the routine design of full bandgap materials in three dimensions for optical wavelengths has proven elusive \cite{john1987strong,yablonovitch1987inhibited, blanco2000large,miguez1997photonic,cubukcu2003electromagnetic,becker2005two,rehammar2010optical}. Recently, disordered and isotropic photonic materials have been suggested as an alternative \cite{reufer2007transport,edagawa2008photonic,florescu2009designer,liew2011photonic,wiersma2013disordered,muller2014silicon}. A hyperuniform structure combined with short range order and an open network architecture is widely considered to be a strong candidate for an optimized photonic material design \cite{florescu2009designer,liew2011photonic,Froufe16PBG}. Moreover the isotropic structure should offer additional advantages such as the possibility to incorporate waveguides with arbitrary bending angles \cite{man2013isotropic,ishizaki2013realization,rinne2008embedded}.  Numerical calculations in two and three dimensions suggest the presence of a full photonic bandgap in the near infrared if the material is made out of silicon with an refractive index $n\simeq3.6$ \cite{florescu2009designer,liew2011photonic}. Previous experimental studies of hyperuniform photonic materials have reported partial and full bandgaps for 2D hyperuniform network structures in the microwave regime \cite{man2013isotropic,man2013photonic}. Recently our group has reported on the fabrication of silicon hyperuniform materials that show a pseudo gap in shortwave infrared \cite{muller2014silicon}. The material however also contained substantial amounts of titania (TiO$_2$, $n \simeq2.4$) somewhat lowering the refractive index of the material. The gold standard to achieve silicon photonic bandgap materials is the silicon double-inversion method, a rather complex multi step process to transfer polymer templates into silicon replica \cite{deubel2004direct,tetreault2006new}. Despite its complexity it has been successfully applied to periodic structures \cite{ho1994photonic,tetreault2006new,hermatschweiler2007fabrication}. Full photonic bandgaps have been reported for silicon woodpile photonic crystals in the shortwave infrared \cite{tetreault2006new} and the near-infrared at telecom wavelengths \cite{staude2010fabrication}. However, the application of this approach to open hyperuniform network structures is even more complicated. In particular retaining the mechanical stability of the network is a challenging task considering the harsh conditions when removing the sacrificial material components after each processing step. Here we report the successful realization of silicon double inversion approach for hyperuniform network structures designed by direct-laser writing into a polymer photo resist \cite{tetreault2006new,hermatschweiler2007fabrication,frolich2013titania}.

\section{Methods}
\subsection{Direct laser writing (DLW)}  Polymeric templates on the mesoscale are then fabricated using a commercially available direct-laser writing (DLW) system (Photonic Professional GT, Nanoscribe GmbH, Germany) in Dip-In configuration. \cite{haberko2013fabrication,PhysRevA.88.043822} The structures are written on infrared transparent CaF$_2$ substrates (Crystan, UK) by dipping an oil-immersion objective (63x, NA=1.4) inside a liquid negative-tone photoresist (IP-DIP, Nanoscribe GmbH, Germany). One should note that the writing process is started in a virtual depth
of approximately $0.5 \mu$ m inside the glass substrate to guarantee a continuous laser writing process along the axial direction
which is necessary to ensure the adhesion of the polymer template
to the substrate. The actual height of the structures is thus reduced by $0.5$ $\mu$ m as well. The highest resolution is achieved by tuning the laser power close to the photopolymerization threshold of the photoresist. The photopolymerized samples are developed in two successive baths of PGMEA (Propylene glycol monomethyl ether acetate) for twice 10 min and consecutively rinsed in a bath of isopropanol for 8 min. Gentle drying is ensured by redirecting a stream of N$_2$ through a bubbler filled with isopropanol onto the sample. Massive square walls are written around each structure to improve the mechanical stability during the development procedure and the post-processing.

\subsection{Silicon double inversion} The DLW produced polymeric HU structures are infiltrated with ZnO using atomic layer deposition (ALD). The depositions is carried out in a commercial ALD reactor (Savannah 100, Cambridge Nanotech, Inc.) operating in exposure mode at a moderate temperature of 110 $^{\circ}$C. Slow heating and cooling ramps are set in order to prevent thermal degradation of the polymer structure. Diethylzinc (Strem Chemicals, Inc., >95 \% purity) and DI (Milli-Q) water were chosen as metal and oxygen sources, respectively. Both precursors are kept in stainless steel reservoirs at room temperature and subsequently introduced by pneumatic valves under a carrier gas flow of 5 sccm N$_2$ into the reactor chamber. For the depositions, pulse durations of 0.015 s are chosen for both the metal precursor and the oxygen source. Each pulse is followed by a dwelling time of 5 s without N$_2$ flow and a purge of 60 s under 20 sccm N$_2$. With these parameters the nominal growth rate is 1.2 \AA\ per cycle. 3000 ALD cycles are applied on the HU structures in order to guarantee complete infiltration. After the ALD process, the ZnO overlayer is removed by plasma etching (PE-100 Series, Plasma Etch, Inc.) with 20 sccm of Ar at a pressure of 0.4 torr (1 torr$\approx$ 133.3 Pa) and a power of 200 W. The etch rate is approximatively 1.3 nm min$^{-1}$. Next, the polymeric fraction of the ZnO-photoresist composite structure is removed via calcination at 500 $^{\circ}$C for >5 hours in a tube furnace (Gero, Type SR(A)). Heating and cooling ramps of 100 $^{\circ}$C per hour are chosen to avoid thermal deterioration and delamination of the structures.
\newline The zinc oxide inverse structures are subsequently infiltrated with amorphous silicon at 480 $^{\circ}$C by thermal chemical vapor deposition. The process is carried out in a custom built reactor operating at a base pressure of 9 Torr by slowly heating the structures with plateaus at 150, 250, 350 and 480 $^{\circ}$C and with a dwelling time between 15-30 min. This allows the structures to thermally stabilize. Disilane gas (Si$_2$H$_6$) (Linde, $\geq 99.998 \%$) is used as precursor. The disilane flow is set to 2 sccm, yielding a growth rate of about 4.6 nm min$^{-1}$. The flow is maintained for 35 min at a pressure of 17 torr in order to completely infiltrate the network structures. The silicon overlayer is partially removed by employing plasma etching with a gas mixture of 10 sccm of Ar and 3 sccm of sulfur hexafluoride (SF$_6$) at a pressure of 0.3 torr and a power of 40 W. The etching rate is 25-30 nm min$^{-1}$. The remaining ZnO is then wet-etched by applying a few drops of aqueous hydrochloric acid (10 vol.$\%$) on the sample. After 1 minute the sample is rinsed with DI water and the procedure is repeated three times until no zinc oxide remains. The sample is dried in a gentle flow of N$_2$. Next, the as-fabricated structures are tempered at 600 $^{\circ}$C for >8 hours in order to transform the amorphous silicon into its polycrystalline phase. This procedure reduces the refractive index slightly but also leads to a lower residual absorption coefficient \cite{becker2005two}. Consecutive plasma etching results in a further reduction of  the silicon overlayer until it was observed to brake off revealing the bare network structure.

\subsection{Electron microscopy} The structures are analyzed by scanning electron microscopy (SEM) (Sirion FEG-XL30 S, FEI) between 5 and 10 kV. Cross-sections are realized by focused ion beam (FIB) milling (Dualbeam NOVA600 Nanolab, FEI) performed by accelerating a Ga ion beam (30 kV) at a current of 3 nA onto the sample for several minutes. A consecutive $"$cleaning$"$ step at 1 nA is used to further clean the etching area.

\subsection{Optical characterization} The optical spectra of the hyperuniform disordered structures are studied using a Fourier transform infrared spectrometer (Bruker Vertex 70, germanium coated KBr beamsplitter) connected to a microscope (Bruker Hyperion 2000, SiC globar light source, liquid N$_2$-cooled InSb detector). The employed objective is a 36x Cassegrain with a numerical aperture of 0.52. Transmittance and reflectance of light incident under a cone of light between 10-30$^{\circ}$ relative to the surface normal are measured. Additional transmittance measurements with a reduced angular spread are performed by tilting the sample with an appropriate holder and shadowing the Cassegrain objective such that a probing illumination under 0-10$^{\circ}$ with respect to normal incidence is achieved. Spectra are normalized by a reference taken in air and on a gold mirror for transmittance and reflectance measurements, respectively. We note that the shifts of the gap wavelength $\lambda_{\text{Gap}}$ observed are well explained by the change in $a$ and the effective refractive index of the material. This excludes that the observed features are due to chemical absorption bands.

\section{Results and discussion}
\begin{figure}
\centering
\includegraphics[width=0.7\textwidth]{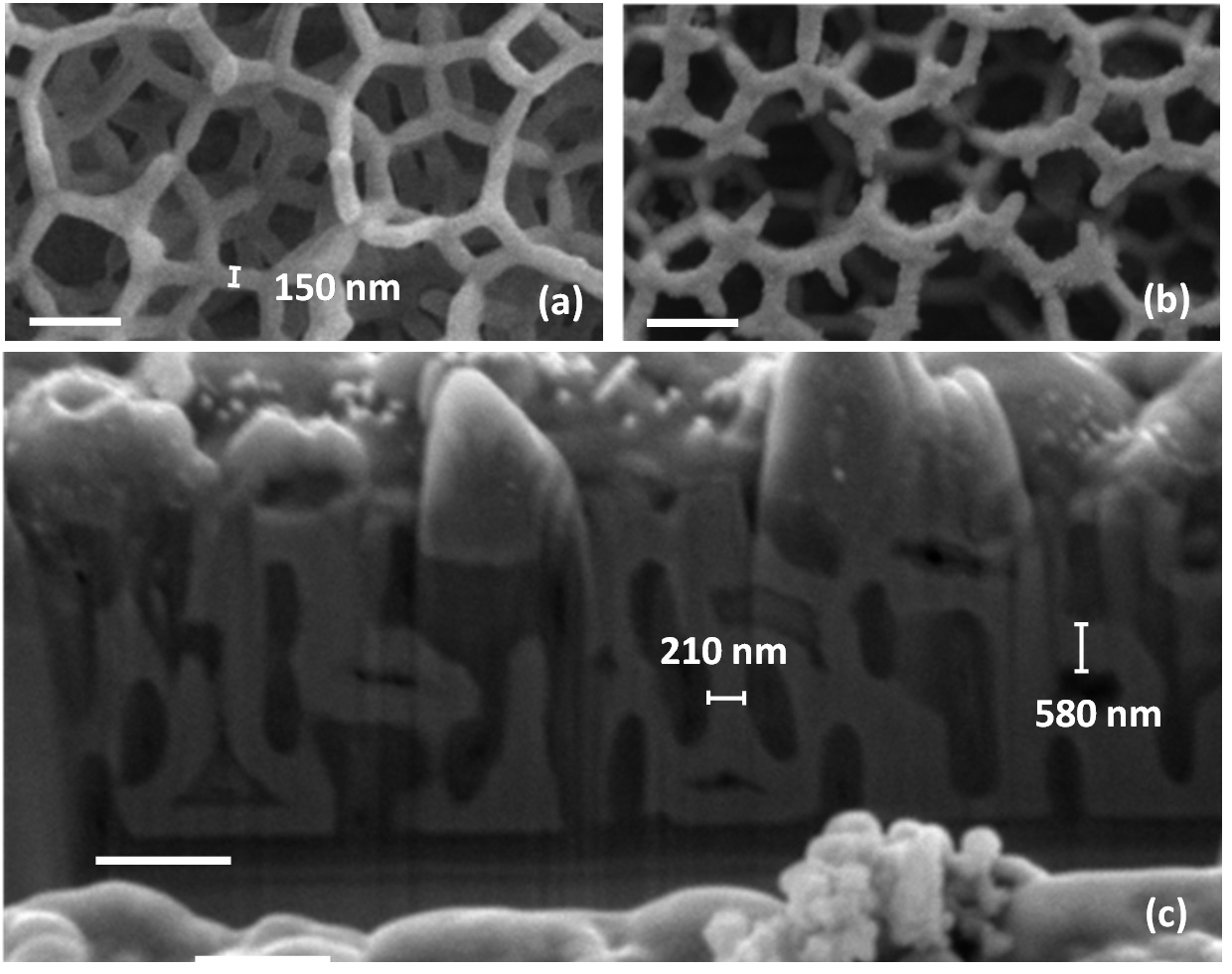}
\caption{Scanning electron microscopy (SEM) of the fabricated hyperuniform disordered structures. (a) Top-view of a polymer structure with $a$=1.82 $\mu$m and height h=5.5 $\mu$m. (b) Image of the back side of a structure composed of poly-crystalline silicon acquired after the silicon double inversion process, revealing that the structure was completely infiltrated during the replication procedure. (c) Close-up view of a focused ion beam (FIB) cross-section of a Si-ZnO composite structure reveals the bulk network structure of solid silicon rods. The dimensions of the rods in-plane is determined to be approximatively 210x580 nm$^2$, resulting in a silicon volume filling fraction of $\phi\approx 0.13$. All scale bars indicate a length of 1 $\mu$m.}
\label{fig:SEM}
\end{figure}

\begin{figure}
\centering
\hspace*{-.5cm}
\includegraphics[width=1\textwidth]{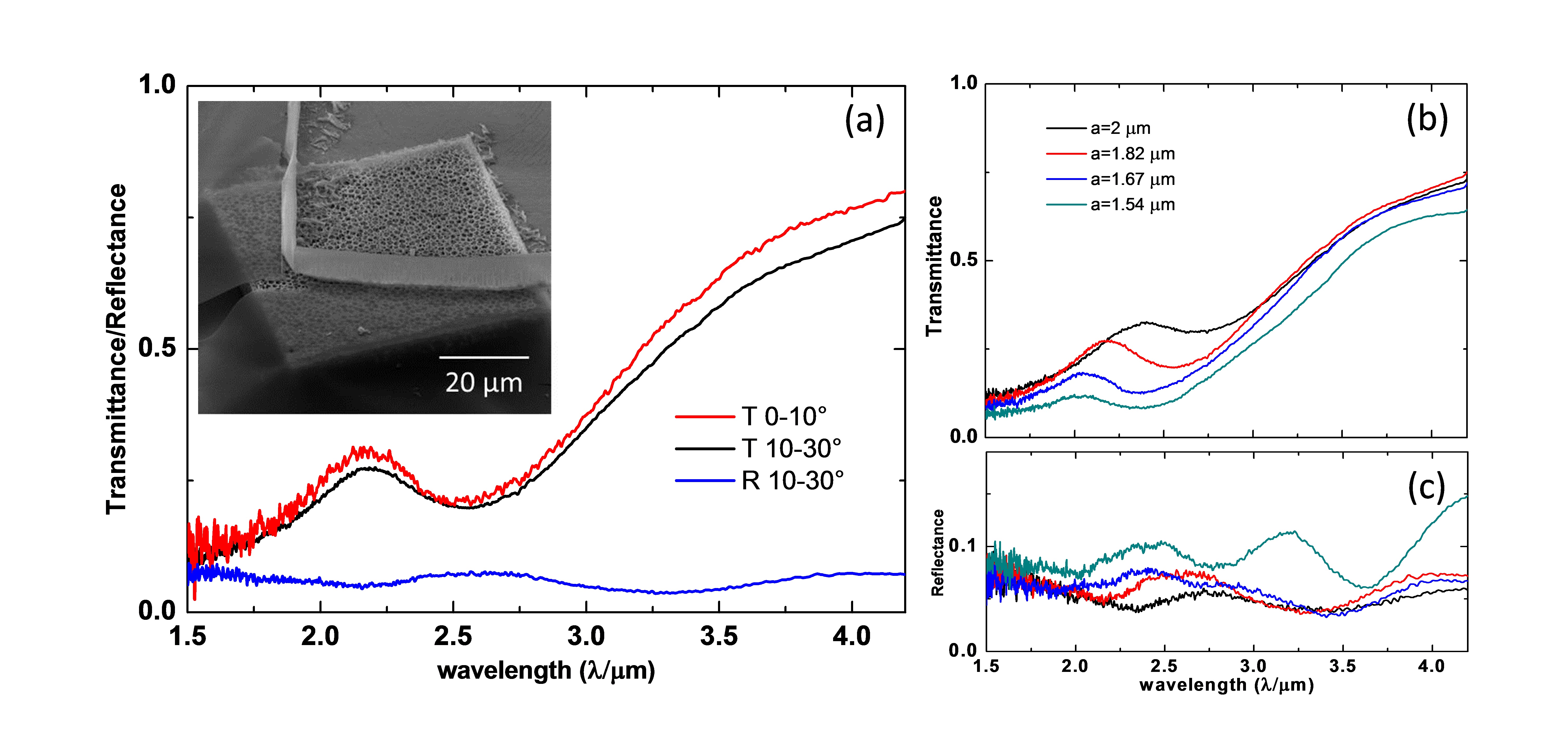}
\caption{Infrared spectroscopy of silicon hyperuniform network structures performed by Fourier transform infrared spectroscopy. (a) Transmittance and reflectance spectra for a hyperuniform structure of $a$=1.82 $\mu$m and height h=5.5 $\mu$m are recorded using a pair of Cassegrain objectives. The resulting probing cone of light possesses an angular spread of 10-30$^{\circ}$ with respect to normal incidence. Additional transmittance measurements are performed with a reduced angular spread of 0-10$^{\circ}$ with respect to the surface normal. Transmission (b) reflectance (c) spectra for h=5.5 $\mu$m and for different values of $a$.  The data displays the blue-shift of the gap wavelength $\lambda_g$  upon reducing the typical structure length scale of the seed pattern $a$= 2, 1.82, 1.67, 1.54 $\mu$m. Inset: Electron micrograph of the silicon material flipped by 180$^{\circ}$ during the final processing step revealing the internal structure of the pure silicon network.}
\label{fig:FTIRgraphs}
\end{figure}

We first fabricate polymeric templates with direct-laser writing (DLW) into a liquid negative-tone photoresist as reported previously in \cite{haberko2013fabrication,PhysRevA.88.043822}. As a seed pattern, we use the center positions of a jammed assembly of spheres \cite{song2008phase}  which is then numerically converted into a three-dimensional hyperuniform (HU) network structure by following the procedure described in reference \cite{haberko2013fabrication}. The average distance between the points is set by the diameter of the jammed spheres and is denoted by $a$. It sets the intrinsic length scale of the structure similar to the lattice constant of a photonic crystal. Consequently, $a$ determines the wavelength of the observed photonic features. The design protocol consists in mapping the seed pattern into tetrahedrons by performing a Delaunay tessellation. Then the centers of mass of the tetrahedrons are connected, resulting in a tetravalent network structure of interconnected rods with the desired structural properties. It has been shown that an approximate relation between the volume filling fraction $\phi$ and the rod diameter <$D$> can be derived as follows \cite{muller2014silicon} : $\phi\approx 3.5($<$D$>$/a)^2$. In the fabrication process the rods acquire an ellipsoidal cross section and an effective value of <$D$> can be calculated by taking the square root of the product between the two axes of the ellipse, Figure \ref{fig:SEM}. The characteristic structural length scale $a$ is varied over a range $a=1.54-2$ $\mu$m. The resulting polymer templates are therefore very open structures with a volume filling fraction of only about 15 \% for a mean rod diameter of <$D$>$\approx$ 400 nm \cite{haberko2013fabrication,PhysRevA.88.043822}. Next, these polymer structures are completely infiltrated with zinc oxide using atomic layer deposition (ALD) at a temperature of 110 $^{\circ}$C, low enough to preserve the integrety of the template. Next the sample is exposed to an Ar plasma in order to remove the excess of ZnO that forms on top of the structure. Then the polymer-ZnO composite material is heated to a temperature of 500 $^{\circ}$C for >5 hours to thermally degrade the polymer. Scanning electron microscopy images confirm the conservation of the network geometry after the thermal treatment. The ZnO inverse structure is infiltrated with amorphous silicon (n$\approx$ 3.6)\cite{janai1979optical} using chemical vapor deposition (CVD) at 480 $^{\circ}$C \cite{hermatschweiler2007fabrication,ishizaki2013realization}. In order to  obtain the positive replica, the ZnO is removed by wet-etching with aqueous hydrochloric acid. At this stage, the photonic features  in the optical spectrum are masked by the pronounced scattering from the Si overlayer (Figure S-2). Further tempering of the structure at 600 $^{\circ}$C for >8 hours transforms the as-deposited amorphous silicon into its brittler poly-crystalline phase. Subsequent Ar-SF$_6$ plasma etching reduces the thickness of the Si overlayer up to a point at which it starts to break off. When doing so the bare network structure becomes accessible, Figure \ref{fig:SEM}(a). We note that this process is rather poorly controlled and the removal of the overlayer remains experimentally challenging. 
We use focussed ion beam (FIB) cutting to access the internal structure of the material. The complete silicon infiltration is confirmed by analyzing cross sections of the FIB cut (Figure \ref{fig:SEM}(c)) as well as by the presence of solid Si rods on the back side of a structure which detached and flipped over during the inversion process (Figure \ref{fig:SEM}(b)). The rods possess an elliptical cross section oriented in plane \cite{PhysRevA.88.043822} with a length of about 210 nm and 580 nm along the short and long axis, respectively. These values corresponding to a mean rod diameter of <$D$>$\approx \sqrt{250\cdot 580}$ nm= 350 nm and a corresponding silicon volume fraction of $\phi\approx$ 0.13. The compositional analysis of the final structure by Energy Dispersive Spectroscopy (Table S-1) indicates the presence of pure silicon.
\begin{figure}
\centering
\includegraphics[width=0.75\textwidth]{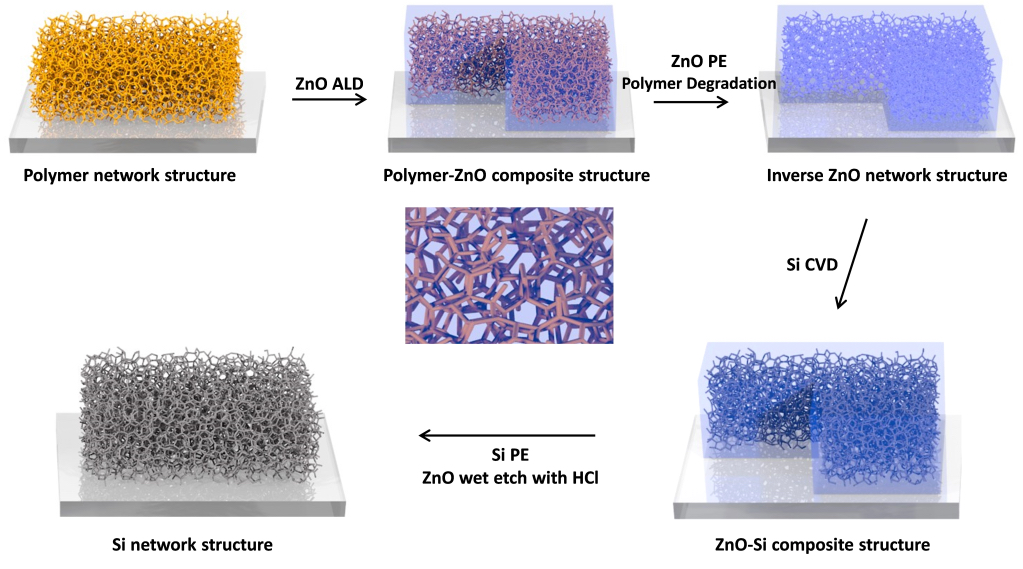}
\caption{Depiction of the double inversion procedure for fabricating silicon replica of a polymer template obtained by direct-laser writing lithography at the mesoscale. (a) Polymeric template of a network structure produced by direct-laser writing (DLW). (b) Completely infiltrated polymer template using ZnO atomic layer deposition (ALD) at a moderate temperature of 110 $^{\circ}$C. (c) Ar plasma etching of the ZnO overlayer and polymer degradation at 500 $^{\circ}$C for > 5 hours. (d) Infiltration of amorphous silicon using chemical vapor deposition (CVD) at 480 $^{\circ}$C. (e) Diluted HCl wet-etching of the remaining ZnO fraction. The Si overlayer is plasma etched using an Ar-SF$_6$ gas mixture. Next, the sample is tempered at 600 $^{\circ}$C for > 8 hours in order to transform the amorphous Si into its poly-crystalline phase.}
\label{fig:SDI}
\end{figure}
\newline The optical spectra of the silicon hyperuniform materials are recorded using a Fourier transform infrared spectrometer coupled to a microscope. We use a pair of Cassegrain lenses with acceptance angles between 10$^{\circ}$ and 30$^{\circ}$ with respect to the surface normal, meaning that a set of reciprocal lattice vectors probe the sample simultaneously. Additional transmittance measurements with an angular spread between 0$^{\circ}$-10$^{\circ}$ along normal incidence were realized by shadowing the Cassegrain condenser objective partially and tilting the sample with a custom made holder similar to the one used in ref. \cite{   staude2010fabrication}. A pronounced gap in transmittance is observed at a central wavelength of $\lambda_{\text{Gap}}\approx$ 2.6 $\mu$m for a structure of height h=5.5 $\mu$m and $a$= 1.82 $\mu$m as shown in Figure \ref{fig:FTIRgraphs} (a). As expected for a nearly isotropic material we find that that the central position $\lambda_{\text{Gap}}$ of the dip as well as its width do not change when measuring at oblique or normal incidence. This is an important observation since disordered materials are isotropic nature and therefore the photonic features are expected to be nearly angular independent\cite{florescu2009designer}. The dip extends from $\lambda\approx$ 2.2-3 $\mu$m while at the same time no strong specular reflectance is observed for the corresponding wavelengths. This means that the light that is not transmitted is diffusely reflected over the whole hemisphere. Residual oscillations in some of the reflectance spectra can be attributed to Fabry-Perrot interference effects.\\
Next we study the evolution of the optical transmission spectra as a function of the typical structure length scale of the seed pattern $a$. To this end a discrete set of network structures is designed and fabricated where the parameter $a$ is reduced from $a$= 2 $\mu$m down to $a$= 1.54 $\mu$m (Figure \ref{fig:FTIRgraphs} (b,c)). Correspondingly the central gap wavelength shifts from $\lambda_{\text{Gap}}\approx$ 2.7 $\mu$m to $\lambda_{\text{Gap}}\approx$ 2.3 $\mu$m. The cross section of the ellipsoidal rods remains unchanged as it is set by direct-laser writing 'pen'. Thus the effective rod diameter <$D$> remains constant as well. This means that by continuously reducing the parameter $a$, the filling fraction increases from $\phi=$0.11 to 0.22. This reduces the blue-shift of $\lambda_{\text{Gap}}$ since the effective refractive index is increased as discussed in more detail in \cite{muller2014silicon}. Numerical calculations on hyperuniform disordered network structures have shown that a complete photonic bandgap appears for filling fractions of $\phi$=0.15-0.4 and for refractive indices n$\geq$ 3 \cite{liew2011photonic}. The optimal filling fraction is predicted to be in the range $\phi$=0.15-0.25. Indeed, it is in this region that we observe experimentally the most pronounced gaps. For $a=1.82$ $\mu$m we find experimentally  $\lambda_{\text{Gap}}\simeq 2.6$ $\mu$m and we estimate a silicon filling fraction of $\phi=0.13$.  For comparison, in the numerical study a central gap wavelength for the silicon network  (n=3.6,$\phi=0.15$) is found at $\lambda_{\text{Gap}}\simeq 1.6 \cdot a=2.9$ $\mu$m \cite{liew2011photonic} which almost quantiatively matches our experimental result. The slight 10\% difference can be attributed to the slightly lower filling fraction and to shrinking effects during the fabrication process. Similar shrinking effects have been reported previously due to the thermal decomposition of the polymer and the high temperature silicon chemical vapor deposition in the fabrication of silicon woodpile photonic crystals \cite{von2010three}.
\newline It would be desirable to reduce the structural length scales even further in order to bring the photonic band gap closer to telecommunication wavelengths. As we have shown here this can only be realized if the cross section of the laser writing 'pen' can be reduced substantially. Otherwise the filling fraction will increase beyond $\phi=0.25$ and the photonic properties are expected to weaken as shown in \cite{liew2011photonic}. Indeed, such a higher resolution can be achieved for example by employing a direct-laser writing scheme based on a 405 nm wavelength diode laser for the fabrication of polymeric templates as reported in \cite{mueller20143d}. We believe the structural length scale $a$ could be reduced to below $a=$1 $\mu$m, thereby shifting the gap to the near-infrared wavelengths around $\lambda_{\text{Gap}}\approx$ 1.3-1.5 $\mu$m.

\section*{Acknowledgements}
This project has been financially supported by the National Research Fund, Luxembourg (project No. 3093332), the Swiss National Science Foundation (projects 132736 and 149867), the National Centre of Competence in Research Bio-Inspired Materials and the Adolphe Merkle Foundation. We thank Bodo Wilts for help with the scanning electron microscopy.

\end{document}